\documentclass[superscriptaddress,reprint,prb]{revtex4-1}
\usepackage{graphicx}
\pdfoutput=1
\usepackage[margin=0.7in]{geometry}
\usepackage{amsmath}
\usepackage{hyperref}
\usepackage{multirow}

\let\hatOrig\hat

\renewcommand{\hat}[1]{\boldsymbol{\mathbf{\hatOrig{#1}}}}
\renewcommand{\Im}{\operatorname{Im}}

\newcommand{\sub}[1]{\ensuremath{_{\textrm{#1}}}} \newcommand{\super}[1]{\ensuremath{^{\textrm{#1}}}}  

\newcommand{\JCAP}{Joint Center for Artificial Photosynthesis, California Institute of Technology, 1200 E. California Blvd, Pasadena CA}
\newcommand{\MSC}{Materials and Process Simulation Center, California Institute of Technology, 1200 E. California Blvd, Pasadena CA}
\newcommand{\RPIMSE}{Department of Materials Science and Engineering, Rensselaer Polytechnic Institute, 110 8\super{th} Street, Troy, NY}
\newcommand{\Watson}{Thomas J. Watson Laboratories of Applied Physics, California Institute of Technology, 1200 E. California Blvd, Pasadena CA}
\newcommand{\NGNEXT}{NG NEXT, 1 Space Park Drive, Redondo Beach CA}
\newcommand{\LBL}{The Molecular Foundry, Lawrence Berkeley National Laboratory, 1 Cyclotron Road, Berkeley CA}

\usepackage[usenames]{color}

\begin{document}

\title{Experimental and \emph{ab initio} ultrafast carrier dynamics in plasmonic nanoparticles}

\author{Ana M. Brown}\affiliation{\Watson}
\author{Ravishankar Sundararaman}\email{sundar@rpi.edu}\affiliation{\JCAP}\affiliation{\RPIMSE}
\author{Prineha Narang}\email{prineha@caltech.edu}\affiliation{\Watson}\affiliation{\JCAP}\affiliation{\NGNEXT}
\author{Adam M. Schwartzberg}\affiliation{\LBL}
\author{William A. Goddard III}\affiliation{\JCAP}\affiliation{\MSC}
\author{Harry A. Atwater}\affiliation{\Watson}\affiliation{\JCAP}

\date{\today}

\begin{abstract}
Ultrafast pump-probe measurements of plasmonic nanostructures probe the non-equilibrium behavior
of excited carriers, which involves several competing effects obscured in typical empirical analyses.
Here we present pump-probe measurements of plasmonic nanoparticles along with
a complete theoretical description based on first-principles calculations
of carrier dynamics and optical response, free of any fitting parameters.
We account for detailed electronic-structure effects in the density of states,
excited carrier distributions, electron-phonon coupling, and dielectric functions which allow us to
avoid effective electron temperature approximations. Using this calculation method, we
obtain excellent quantitative agreement with spectral and temporal features in transient-absorption measurements.
In both our experiments and calculations, we identify the two major contributions of the initial response
with distinct signatures: short-lived highly non-thermal excited carriers and longer-lived thermalizing carriers.
\end{abstract}

\maketitle

Plasmonic hot carriers provide tremendous opportunities for combining efficient light capture with energy
conversion\cite{Brongersma:2015fk, Atwater:2010ys, Leenheer:2014sw, Catchpole2012, Prineha:2016fk}
and catalysis\cite{Linic:2013tu, Mukherjee:2012uq} at the nano scale.\cite{Clavero:2014vn, Linic:2015kx, Moskovits:2015sf}
The microscopic mechanisms in plasmon decays across various energy, length
and time scales are still a subject of considerable debate, as seen in recent
experimental\cite{Harutyunyan:2015nx, Zavelani-Rossi:2015kx} and theoretical
literature.\cite{Govorov:2013fk, Babicheva:2015uq, Govorov:2014zh, PhysRevB.85.245423}
The decay of surface plasmons generates hot carriers through several mechanisms
including direct interband transitions, phonon-assisted intraband transitions and
geometry-assisted intraband transitions, as we have shown in previous work.\cite{PhononAssisted, NatCom}

\begin{figure}
\includegraphics[width=\columnwidth]{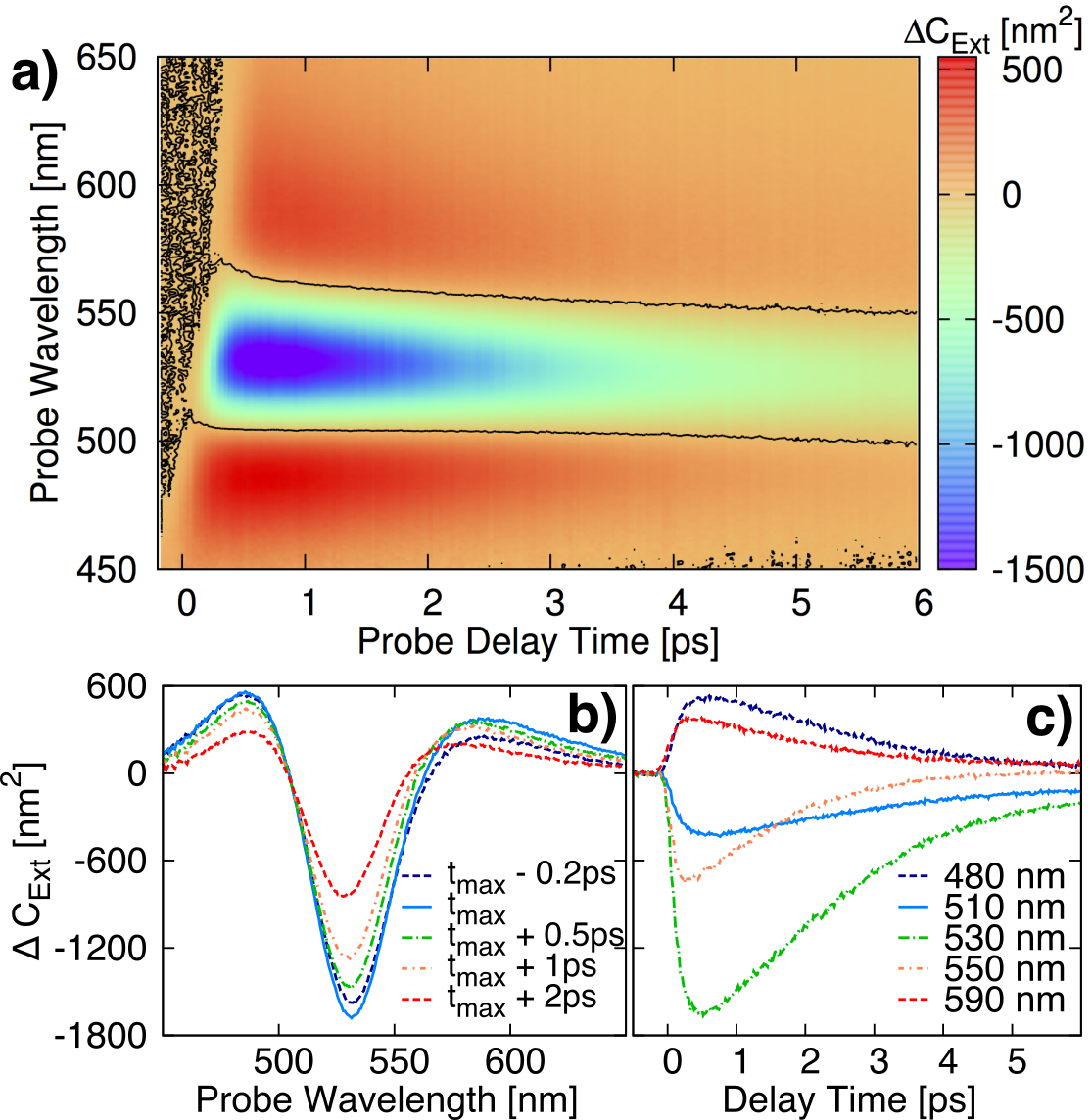}
\caption{
(a) Map of the differential extinction cross section of colloidal gold nanoparticles
as a function of pump-probe delay time and probe wavelength
for a pump pulse of 68 $\mu \text{J}/\text{cm}^2$ energy density at 380 nm.
At time 0, the pump pulse excites the sample.
As the electrons thermalize internally, extinction near the
absorption peak (533 nm) decreases (negative signal)
while extinction in the wings to either side
of the absorption peak increases (positive signal).
After $\sim$ 700 fs, the electrons began to
thermalize with the lattice and the differential extinction decays.
A contour line is drawn in black at zero extinction change.
Differential extinction (b) as a function of
probe wavelength at a set of times relative
to the pump-probe delay time with maximum signal,  $t\sub{max} = 700$ fs; and
(c) as a function of pump-probe delay time at various probe wavelengths.
\label{fig:deltaCextMap}}
\end{figure}

Dynamics of hot carriers are typically studied via ultrafast pump-probe measurements
of plasmonic nanostructures
using a high-intensity laser pulse to excite a large number
of electrons and measure the optical response as a function
of time using a delayed probe pulse.\cite{
Anisimov, DelFatti, Elsayed1987, Elsayed-Ali, Giri, Hartland, Kaganov, Harutyunyan:2015nx}
Various studies have taken advantage of this technique
to investigate electron-electron scattering,
electron-phonon coupling, and electronic transport.\cite{DelFatti,Sun,
Groeneveld1995,Elsayed-Ali,knoesel1998ultrafast, aeschlimann1997ultrafast,
hohlfeld1997nonequilibrium, brorson1987femtosecond, PhysRevB.75.155426}
Fig.~\ref{fig:deltaCextMap} shows a representative map
of the differential extinction cross section
as a function of pump-probe delay time and probe wavelength.
With an increase in electron temperature, the real part of the
dielectric function near the resonant frequency becomes more negative,
while the imaginary part increases.\cite{TAparameters}
This causes the resonance to broaden and blue shift at short times
as the electron temperature rises rapidly, and then to narrow
and shift back over longer times as electrons cool down,
consistent with previous observations.\cite{PhysRevB.80.245420}
Taking a slice of the map at one probe wavelength reveals the temporal behavior 
of the electron relaxation (Fig.~\ref{fig:deltaCextMap}(b)) whereas a slice of 
the map at one time gives the spectral response, as shown in 
Fig.~\ref{fig:deltaCextMap}(a) for a set of times relative to the delay time 
with maximum signal, $t_{max} = 700$ fs.

Conventional analyses of pump-probe measurements invoke a `two-temperature model'
that tracks the time dependence (optionally the spatial variation)
of separate electron and lattice temperatures, $T_e$ and $T_l$ respectively,
which implicitly neglects non-equilibrium effects of the electrons.
Recent literature has focused on the contributions of thermalized and nonthermalized electrons
to the optical signal in pump-probe measurements using free-electron-like theoretical models
to interpret optical signatures.\cite{DelFatti,Shen,Sun,Della,Voisin,giri2015transient}
However, these models invariably require empirical parameters for both the dynamics
and response of the electrons, making unambiguous interpretation of experiments challenging.
This \emph{Letter} quantitatively identifies non-equilibrium ultrafast dynamics of electrons,
combining experimental measurements and parameter-free \emph{ab initio} predictions
of the excitation and relaxation dynamics of hot carriers in plasmonic metals
across timescales ranging from 10~fs--10~ps.
Note that, while metal thin films or single crystals would provide a `cleaner'
experimental system in general, we focus on nanoparticles here
because they enable an important simplification: electron distributions are
constant in space over the length scale of these particles, allowing us
to treat temporal dynamics and optical response in greater detail.
(See supplementary information.)

A theoretical description of pump-probe measurements of hot carrier dynamics
in plasmonic systems involves two major ingredients: 
i) The optical response of the metal (and its environment)
determines the excitation of carriers by the pump
as well as the subsequent signal measured by the probe pulse.
ii) The dynamics of the excited carriers,
including electron-electron and electron-phonon scattering,
determines the time dependence of the probe signal.
We previously presented\cite{TAparameters} \emph{ab initio}
theory and predictions for both the optical response
and the dynamics within a two temperature model,
where the electrons are assumed to be in internal equilibrium
albeit at a different temperature from the lattice.
Below, we treat the response and relaxation
of non-thermal electron distributions from first principles,
without assuming an effective electron temperature at any stage.

\begin{figure}
\includegraphics[width=0.8\columnwidth]{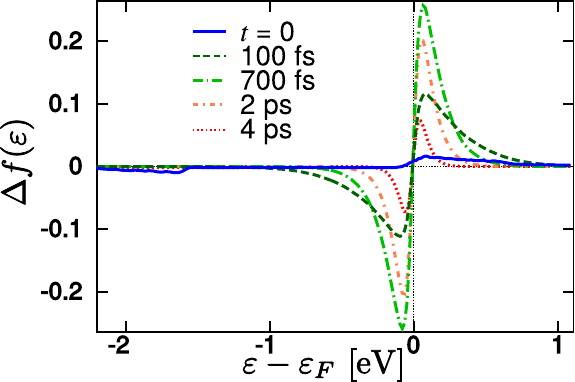}
\caption{Difference of the predicted time-dependent electron distribution 
from the Fermi distribution at 300 K, induced by a pump pulse
at 560 nm with intensity of 110 $\mu \text{J}/\text{cm}^2$.
Starting from the carrier distribution excited by plasmon decay at $t=0$,
electron-electron scattering concentrates the distribution near the Fermi level
with the peak optical signal at $\sim$700~fs,
followed by a return to the ambient-temperature Fermi distribution
and a decay of the optical signal due to electron-phonon scattering.
\label{fig:DistributionChanges}}
\end{figure}

For the optical response, we calculate the imaginary part
of the dielectric function $\Im\epsilon(\omega)$
accounting for direct interband transitions,
phonon-assisted intraband transitions
and the Drude (resistive) response,
and calculate the real part using the Kramers-Kronig relations.
Specifically, we start with density-functional theory calculations
of electron and phonon states as well as electron-photon
and electron-phonon matrix elements using the JDFTx code,\cite{JDFTx}
convert them to an \emph{ab initio} tight-binding model using Wannier functions,\cite{MLWFmetal}
and use Fermi Golden rule and linearized Boltzmann equation
for the transitions and Drude contributions respectively.
The theory and computational details for calculating $\epsilon(\omega)$
are presented in detail in Refs.~\citenum{PhononAssisted} and
\citenum{TAparameters}, and we do not repeat them here.
All these expressions are directly in terms of the electron
occupation function $f(\varepsilon)$, and we can straightforwardly
incorporate an arbitrary non-thermal electron distribution
instead of Fermi functions.
These non-thermal distributions differ from the thermal Fermi distributions
by sharp distributions of photo-excited electrons and holes that dissipate with time
due to scattering, as shown in Fig.~\ref{fig:DistributionChanges} and discussed below.

We use the \emph{ab initio} metal dielectric function for
calculating the initial carrier distribution as well as the probed response.
The initial carrier distribution following the pump pulse is given by
\begin{equation}
f(\varepsilon, t=0) = f_0(\varepsilon) + U\frac{P(\varepsilon, \hbar\omega)}{g(\varepsilon)}
\label{eqn:distribution}
\end{equation}
where $f_0$ is the Fermi distribution at ambient temperature,
$U$ is the pump pulse energy absorbed per unit volume,
$g(\varepsilon)$ is the electronic density of states,\cite{TAparameters}
and $P(\varepsilon, \hbar\omega)$ is the energy distribution of carriers
excited by a photon of energy $\hbar\omega$.\cite{PhononAssisted}
We then evolve the carrier distributions and lattice temperature in time
to calculate $f(\varepsilon,t)$ and $T_l(t)$ as described next.
From those, we calculate the variation of the metal dielectric function $\epsilon(\omega,t)$,
and in turn, the extinction cross section using Mie theory.\cite{MieTheory,MieMatlab}
To minimize systematic errors between theory and experiment,
we add the \emph{ab initio} prediction for the change in
the dielectric function from ambient temperature,\cite{TAparameters}
to the experimental dielectric functions from ellipsometry.\cite{Palik1985}

We calculate the time evolution of the carrier distributions
using the nonlinear Boltzmann equation
\begin{equation}
\frac{d}{dt}f(\varepsilon,t)
= \Gamma\sub{e-e}[f](\varepsilon)
+ \Gamma\sub{e-ph}[f,T_l](\varepsilon),
\label{eqn:boltzmann}
\end{equation}
where  $\Gamma\sub{e-e}$ and $\Gamma\sub{e-ph}$, respectively,
are the contributions due to electron-electron and
electron-phonon interactions to the collision integral.
For simplicity, we assume that the phonons remain thermal
at an effective temperature $T_l(t)$ and calculate the
time evolution of the lattice temperature using energy balance,
$-C_l(T_l) (dT_l/dt) = (dE/dt)|\sub{e-ph}$,
where the term on the right corresponds to the rate of energy
transfer from the lattice to the electrons due to $\Gamma\sub{e-ph}$,
and $C_l$ is the \emph{ab initio} lattice heat capacity.\cite{TAparameters}

The \emph{ab initio} collision integrals are extremely computationally expensive
to calculate repeatedly to directly solve (\ref{eqn:boltzmann}).
We therefore use simpler models for the collision integrals
parametrized using \emph{ab initio} calculations.
For electron-electron scattering in plasmonic metals,
the calculated electron lifetimes exhibit the inverse quadratic energy dependence
$\tau^{-1}(\varepsilon) \approx (D_e/\hbar) (\varepsilon - \varepsilon_F)^2$
characteristic of free electron models within Fermi liquid theory.\cite{PhononAssisted}
We therefore use the free-electron collision integral,\cite{Sun,mueller2013relaxation,DelFatti}
\begin{multline}
\Gamma\sub{e-e}[f](\varepsilon) = \frac{2 D_e}{\hbar}
	\int d\varepsilon_1  d\varepsilon_2 d\varepsilon_3
	\frac{g(\varepsilon_1) g(\varepsilon_2) g(\varepsilon_3)}{g^3(\varepsilon_F)}\\
	\times \delta(\varepsilon + \varepsilon_1 - \varepsilon_2 - \varepsilon_3)
	\big[ f(\varepsilon_2) f(\varepsilon_3) (1-f(\varepsilon)) (1-f(\varepsilon_1)) \\
	- f(\varepsilon) f(\varepsilon_1)(1-f(\varepsilon_2))(1-f(\varepsilon_3)) \big]
\label{eqn:eeCollision}
\end{multline}
with the constant of proportionality $D_e$ extracted from
\emph{ab initio} calculations of electron lifetimes.\cite{TAparameters}
In doing so, we neglect variation of the electron-electron scattering rate
between states with different momenta at the same energy, which is an excellent
approximation for gold where this variation is $\sim 10\%$ for energies within 5~eV of the Fermi level.\cite{PhononAssisted}
For electron-phonon scattering, assuming that phonon energies are negligible
on the electronic energy scale (an excellent approximation for
optical frequency excitations in metals),
we can simplify the electron-phonon collision integral to
\begin{multline}
\Gamma\sub{e-ph}[f,T_l](\varepsilon) = \\
\frac{1}{g(\varepsilon)} \frac{\partial}{\partial \varepsilon} \left[
	H (\varepsilon)
	\left( f (\varepsilon) (1 - f(\varepsilon)) + k_B T_l \frac{\partial f}{\partial \varepsilon} \right)
\right],
\label{eqn:Gammaeph}
\end{multline}
where $H(\varepsilon)$ is an energy-resolved electron-phonon coupling strength
calculated from \emph{ab initio} electron-phonon matrix elements.\cite{TAparameters}
(See Supporting Information for details, derivations and
plots as well as numerical tabulation of $H(\varepsilon)$ for four
commonly used plasmonic metals: the noble metals and aluminum.)

In our experiments, we use 
an ultrafast transient absorption system with a tunable pump
and white light probe probe
to measure the extinction of Au colloids in solution
as a function of pump-probe delay time
and probe wavelength. The laser system consists of a 
regeneratively amplified Ti:sapphire oscillator
(Coherent Libra),
which delivers 1mJ pulse energies centered at 800 nm with a 1 kHz repetition rate.
The pulse duration of the amplified pulse is approximately 50 fs.
The laser output is split by an optical wedge to produce the pump and probe beams and
the pump beam wavelength is tuned using a coherent OperA OPA.
The probe beam is focused onto a sapphire plate to generate a white-light continuum probe.
The time-resolved differential extinction spectra are collected
with a commercial Helios absorption spectrometer (Ultrafast Systems LLC).
The temporal behavior is monitored by increasing the path length
of the probe pulse and delaying it with respect to the pump pulse
with a linear translation stage capable of step sizes as small as 7 fs.
Our sample is a solution of 60-nm-diameter Au colloids in water
with a concentration of $2.6 \times 10^{10}$ particles per milliliter
(BBI International, EM.GC60, OD1.2)
in a quartz cuvette with a 2 mm path length.

The initial excitation by the pump pulse generates an electron distribution
that is far from equilibrium, for which temperature is not well-defined.
Our \emph{ab initio} predictions of the carrier distribution at $t=0$
in Fig.~\ref{fig:DistributionChanges} exhibits high-energy holes in the
$d$-bands of gold and lower energy electrons near the Fermi level.
These highly non-thermal carriers rapidly decay within 100~fs,
resulting in carriers closer to the Fermi level
which thermalize via electron-electron scattering in several 100~fs,
reaching a peak higher-temperature thermal distribution at $\sim 700$~fs
in the example shown in Fig.~\ref{fig:DistributionChanges}.
These thermalized carriers then lose energy to the lattice
via electron-phonon scattering over several picoseconds.

The conventional two-temperature analysis is only valid in that last phase
of signal decay (beyond 1~ps) once the electrons have thermalized.
The initial response additionally includes contributions from
short-lived highly non-thermal carriers excited initially,
that become particularly important at low pump powers when
smaller temperature changes limit the thermal contribution.
Higher energy non-thermal carriers exhibit faster rise and decay times
than the thermal carriers closer to the Fermi level,\cite{Sun,Shen}
due to higher electron-electron scattering rates.
Their response also spans a greater range in probe wavelength
compared to thermal electrons which primarily affect only the resonant
$d$-band to Fermi level transition.\cite{Sun,Hohlfeld,TAparameters}
Combining \emph{ab initio} predictions and experimental measurements
of 60-nm colloidal gold solutions, we quantitatively identify
these signatures of thermal and non-thermal electrons,
first as a function of pump power and then as a function of probe wavelength.

\begin{figure}
\includegraphics[width=0.9\columnwidth]{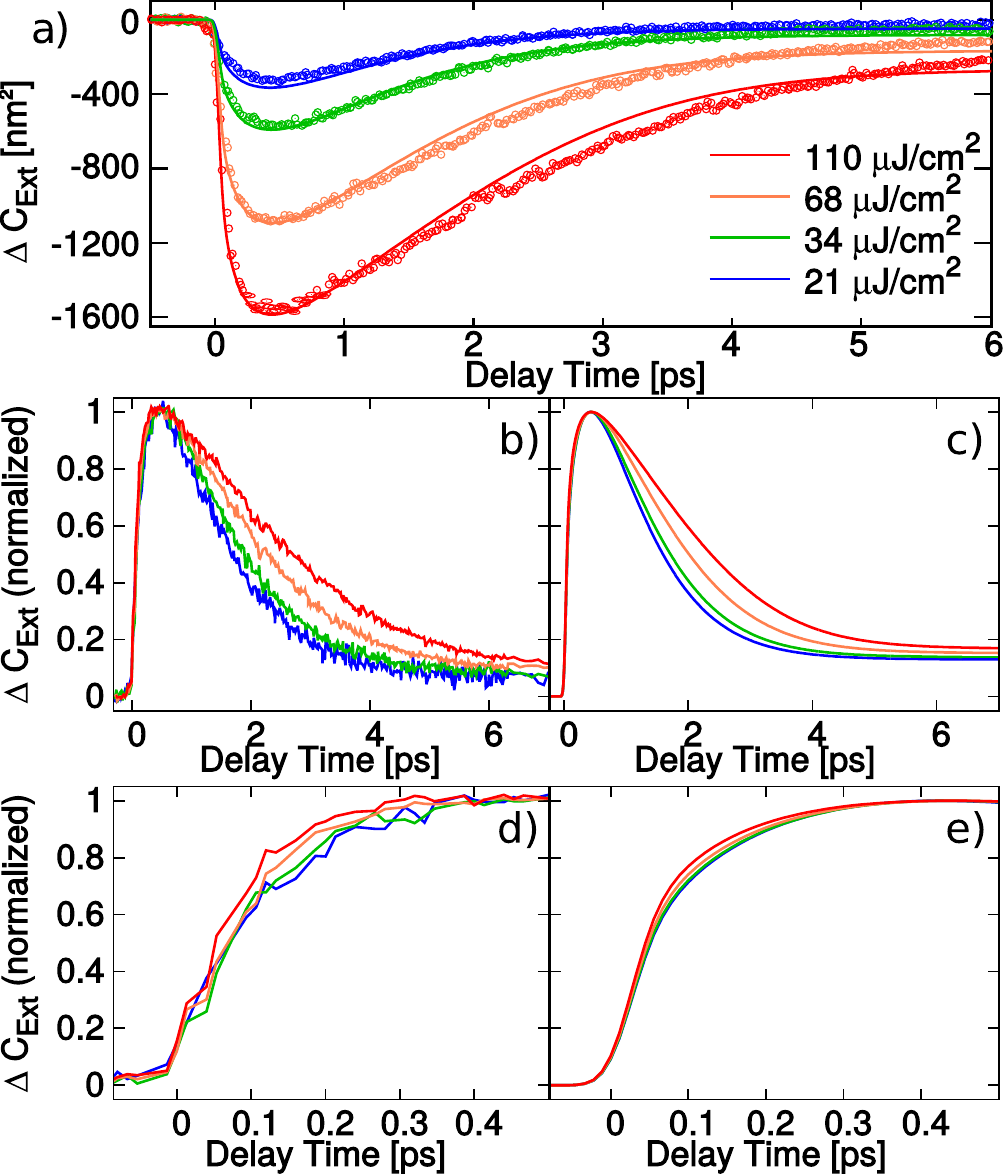}
\caption{
Comparison of measured and predicted differential cross sections at 530~nm probe wavelength
for pump excitation at 560~nm with intensities of 21, 34, 68, and 110~$\mu$J/cm$^2$ as a function of time.
Part (a) compares absolute measurements (circles) and calculated values (solid lines)
of the differential cross-section, while the remaining parts normalized by the peak value:
(b) and (c) show measurements and predictions respectively over the full time range,
while (d) and (e) focus on the initial rise period.
Increased pump power generates more initial carriers, which equilibrate faster
(shorter rise time) to a higher electron temperature (larger signal amplitude),
which subsequently relaxes more slowly due to increased electron heat capacity.
The \emph{ab initio} predictions quantitatively match all these features of the measurements.
\label{fig:PumpPowerThermal}}
\end{figure}

Fig.~\ref{fig:PumpPowerThermal}(a) first shows that our \emph{ab initio} predictions of
electron dynamics and optical response quantitatively capture the \emph{absolute}
extinction cross section as a function of time for various pump pulse energies.
Note that the agreement is uniformly within 10\%, which is the level
of accuracy that can be expected for parameter-free DFT predictions,
given that the first-principles band structures are accurate to
0.1 -- 0.2~eV and optical matrix elements are accurate to 10 -- 20\%,
with the larger errors for localized $d$ electrons.\cite{NatCom}
We then examine the cross section time dependence normalized
by peak values to more clearly observe the changes in rise and decay time scales.

Decay of the measured signal is because of energy transfer
from electrons to the lattice via electron-phonon scattering.
At higher pump pulse energies, the electrons thermalize to a higher temperature.
For $T_e < 2000$~K, the electron heat capacity increases linearly with temperature,
whereas the electron-phonon coupling strength does not appreciably change
with electron temperature.\cite{DelFatti,TAparameters}
Therefore, the electron temperature, and correspondingly the measured
probe signal, decays more slowly at higher pump powers
as shown in Fig.~\ref{fig:PumpPowerThermal}(b,c).
Again, we find quantitative agreement between the measurements
and \emph{ab initio} predictions with no empirical parameters.

Rise of the measured signal arises from electron-electron scattering
which transfers the energy from few excited non-thermal electrons
to several thermalizing electrons closer to the Fermi level.
Higher power pump pulses generate a greater number of initial
non-thermal carriers, requiring fewer electron-electron collisions
to raise the temperature of the background of thermal carriers.
Additionally, the electron-electron collision rate increases
with temperature because of increased phase space for scattering.\cite{DelFatti}
Both these effects lead to a faster rise time at higher pump powers,
as seen in the measurements shown in Fig.~\ref{fig:PumpPowerThermal}(d),
as well as in the \emph{ab initio} predictions shown in
Fig.~\ref{fig:PumpPowerThermal}(e), once again in quantitative agreement.

\begin{figure}
\includegraphics[width=\columnwidth]{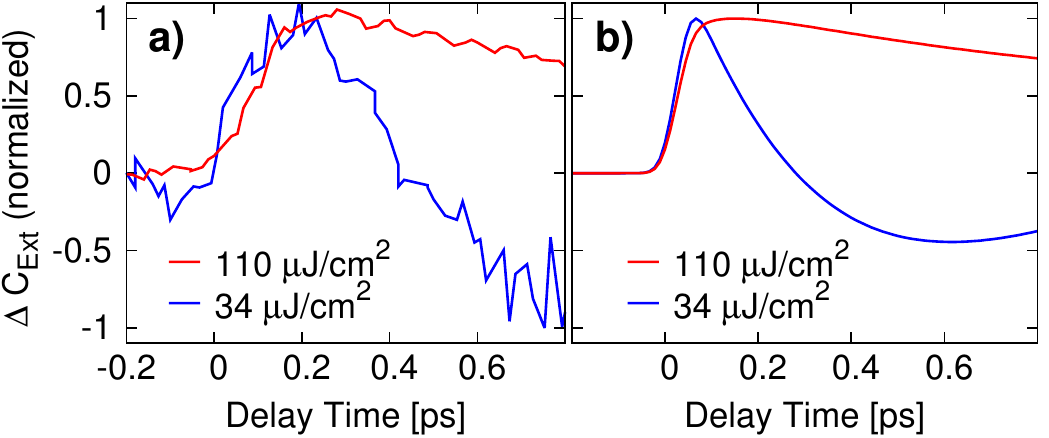}
\caption{
(a) Measured and (b) calculated differential cross sections normalized
by peak value for 380~nm pump pulse with 34 and 110~$\mu$J/cm\super{2}
intensities, monitored at 560~nm probe wavelength.
Contributions from the nonthermal electrons dominate at
lower pump power, resulting in a fast signal rise and decay.
(Correspondingly smaller signals cause the higher relative noise
in the measurements shown in (a).)
\label{fig:PumpPowerNonThermal}}
\end{figure}

Next, we examine the variation of the ratio of
thermal and non-thermal electron contributions with pump power.
Fig.~\ref{fig:PumpPowerNonThermal} shows the sub-picosecond variation
of measured response for two different pump powers,
but now with a pump wavelength of 380~nm with a higher energy photon
that excites non-thermal carriers further from the Fermi level.
Additionally, the probe wavelength of 560~nm is far from
the interband resonance at $\sim 520$~nm, so that the
thermal electrons contribute less to the measured response.
The response has a slow rise and decay time for the higher pump power,
as observed previously in cases where thermal electrons dominate.
However for the lower pump power, the thermal contribution is smaller
making the non-thermal contribution relatively more important,
resulting in a faster rise and decay time.
Once again, the measurements and \emph{ab initio} calculations, which
include all these effects implicitly, are in quantitative agreement.

\begin{figure}
\includegraphics[width=\columnwidth]{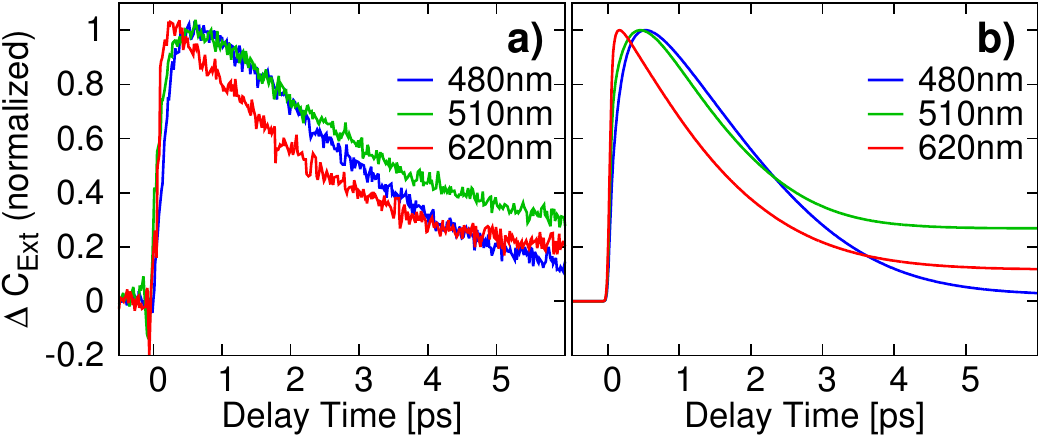}
\caption{
(a) Measured and (b) calculated differential cross sections for 
560 nm pump pulse with 110 $\mu \text{J}/\text{cm}^2$ intensity,
normalized by peak value, for probe wavelengths of 480, 510, and 620 nm.
Rise and decay are faster for probe wavelengths far from the interband
resonance at 530~nm, where non-thermal effects are relatively more important.
\label{fig:ProbeWavelength}}
\end{figure}

Finally, we examine the variation of the temporal signatures with probe wavelength.
Thermalized electrons in noble metals predominantly contribute near the
resonant $d\rightarrow s$ transitions, and therefore non-thermal signatures
become relatively more important at probe wavelengths far from these resonances.
Fig.~\ref{fig:ProbeWavelength}(a) indeed shows a faster rise and decay
due to non-thermal electrons for a probe wavelength of 620~nm,
compared to that at 510~nm which is near the interband resonance (530~nm).
Capturing the wavelength dependence of the dielectric function 
in simple theoretical models\cite{Sun} is challenging because
it involves simultaneous contributions from a continuum
of electronic transitions with varying matrix elements.
Our \emph{ab initio} calculations (Fig.~\ref{fig:ProbeWavelength}(b))
implicitly account for all these transitions and are therefore
able to match both the spectral and temporal features of
the measurements, with no empirical parameters.

To conclude, by combining the first principles calculations of carrier dynamics and optical response
this \emph{Letter} presents a complete theoretical description of pump-probe measurements,
free of any fitting parameters that are typical in previous analyses.\cite{JAP2015Giri,Norris,ParabolicBandModel,Shen}
The theory here accounts for detailed energy distributions of excited carriers (Fig.~\ref{fig:DistributionChanges})
instead of assuming flat distributions,\cite{Della,mueller2013relaxation,Voisin}
and accounts for electronic-structure effects in the density of states,
electron-phonon coupling and dielectric functions beyond the empirical
free-electron or parabolic band models previously employed.\cite{
JAP2015Giri,Norris,Lin,Wang,mueller2013relaxation,Sun,Rethfeld,ParabolicBandModel,DelFatti,Voisin}
This framework, by leveraging Wannier interpolation of electron-phonon matrix elements,
enables quantitative predictions, while avoiding the empiricism that could
hide cancellation of errors or obscure physical interpretation of experimental data.
For example, we clearly identified the temporal and spectral signatures of
short-lived highly nonthermal initial carriers and the longer-lived
thermalizing carriers near the Fermi level in plasmonic nanoparticles.
By demonstrating the predictive capabilities of our theory for metal nanoparticles,
we open up the field for similar studies in other materials\cite{Alkauskas:2016rm} where fits
are not necessarily possible or even reliable eg. semiconductor plasmonics,
and where \emph{ab initio} theory of ultrafast dynamics will be indispensable.

\textbf{Acknowledgements}:
This material is based upon work performed by the Joint Center for Artificial Photosynthesis,
a DOE Energy Innovation Hub, supported through the Office of Science
of the U.S. Department of Energy under Award Number DE-SC0004993.
Work at the Molecular Foundry was supported by the Office of Science,
Office of Basic Energy Sciences, of the U.S. Department of Energy under Contract No. DE-AC02-05CH11231.
The authors acknowledge support from NG NEXT
at Northrop Grumman Corporation. Calculations in this work used the National Energy 
Research Scientific Computing Center,
a DOE Office of Science User Facility supported by the Office of Science of
the U.S. Department of Energy under Contract No. DE-AC02-05CH11231.
P. N. is supported by a National Science Foundation Graduate Research Fellowship
and by the Resnick Sustainability Institute.
A. B. is supported by a National Science Foundation Graduate Research Fellowship,
a Link Foundation Energy Fellowship, and the DOE `Light-Material Interactions
in Energy Conversion' Energy Frontier Research Center (DE-SC0001293).

\bibliographystyle{apsrev4-1}
\makeatletter{} 

\end{document}